\documentclass[twocolumn,preprintnumbers,amsmath,amssymb]{revtex4}
\usepackage{amsmath,amssymb,latexsym,epsfig,graphicx,epstopdf,epsf,pdfpages,dcolumn}
\begin{document}

\title{Quasi-conical quantum dot: electron states and quantum transitions}
\author{Eduard M Kazaryan${}^1$}
\author{Lyudvig S. Petrosyan${}^{1,2}$}
\altaffiliation{email:petrosyan.lyudvig@gmail.com}
\author{Vanik A Shahnazaryan${}^{1,3}$}
\altaffiliation{email:vanikshahnazaryan@gmail.com}
\author{Hayk A Sarkisyan${}^{1,4}$}
\altaffiliation{email:shayk@ysu.am}
\affiliation{${}^1$Russian-Armenian (Slavonic) University, 0051 Yerevan, Armenia}
\affiliation{${}^2$Yerevan State Medical University, 0025 Yerevan, Armenia}
\affiliation{${}^3$Science Institute, University of Iceland, Dunhagi-3, IS-107, Reykjavik, Iceland}
\affiliation{${}^4$Yerevan State University, 0025 Yerevan, Armenia}

\begin{abstract}

The exactly solvable model of quasi-conical quantum dot, having a form of spherical sector is proposed. Due to the specific symmetry of the problem the separation of variables in spherical coordinates is possible in the one-electron Schrodinger equation. Analytical expressions for wave function and energy spectrum are obtained. It is shown that at small values of the stretch angle of spherical sector the problem reduced to the conical QD problem. The comparison with previously performed works showed good agreement of results. As an application of the obtained results, the quantum transitions in the system are considered.

\textbf{Keywords}: Conical quantum dot, electron states, quantum transitions

\pacs{73.21.La, 73.22.Dj}
\end{abstract}

\maketitle

\section{Introduction}

Modern technologies of zero-dimensional semiconductor structures growth allow to realize quantum dots (QD) of different shapes and geometry [1]. The presence of size quantization in all three directions makes the energy spectrum of charge carriers atom-like and strongly dependent on the geometry and linear dimensions of the sample [2]. The knowledge of the characteristics of the band structure of the QD makes it possible to reveal the features of physical parameters characterizing the investigated sample. This fact allows one to consider QDs as promising candidates for element base of the semiconductor devices of new generation. It is clear that the more geometrical parameters characterize QD, the more flexibly one can control its energy spectrum. For instance, in the case of spherical QD [3-5] the only geometrical parameter is its radius. In the case of cylindrical QD [6-10] there are two such parameters: the cross-sectional radius and height of the cylinder. For the spherical quantum layer [11,12] the geometrical parameters are the internal and external radiuses. The cylindrical quantum layer already has three geometric parameters: the height of the cylinder, and its inner and outer radiuses [13-15]. The above examples have either spherical or cylindrical geometry. This allows to spend quite a substantial mathematical analysis of the relevant solutions of Schrodinger equations, as well as the nature of the energy spectrum of electron states [16]. However it is possible to realize QDs having more complex geometry, the analytical description of which is not so trivial. The examples of such systems are pyramidal and conical QDs which can be obtained e.g. using a self-organizing growth method [17, 18]. In the mentioned systems in addition to hardships connected with the description of the complex geometry, it is necessary to model the confinement potential of QD, taking into account the strain effects on the interface QD - environment [19, 20]. These circumstances force to describe the physical properties of pyramidal and conical QDs numerically. In particular, in Ref. [19] the strain distribution in and around pyramidal InAs/GaAs QDs on a thin wetting layer is simulated numerically. In its turn, for QD having a conical shape the authors [21] proposed a general numerical methodology that can be applied to both isolated and coupled quantum dots and presented results for the simple one-band model case. In Ref. [22] the authors studied electronic states and optical transitions in a conical QD on the basis of the finite element method. In Ref. [23] carrier capture rates into cone- and truncated-cone-shaped quantum dots mediated by Auger processes are calculated. It is demonstrated that the capture rates strongly depend on both dot size and shape. In Ref. [24] equilibrium composition maps in InGaAs/GaAs conic strained quantum dots, using the finite element method and quadratic programming optimization method are presented.

On the other hand, if one consider the confinement potential of pyramidal or conical QD relatively simple, assuming that the electron is in a impenetrable QD with zero potential energy inside, it is possible to obtain a number of analytical results for the wave functions and the energy spectrum of QD [25]. It should be mentioned, that on the basis of the results obtained in Ref. [25], the authors [26] have studied the effects of hydrostatic pressure, temperature and impurity position on the donor binding energy of a pyramid QD.  Note that a QD having a shape of spherical sector also can be approximated to conical quantum dot (see Fig. 1). The geometry of such system allows one to make the separation of variables in Schrodinger equation without any approximations and obtain an analytical expression for electron wave function. The aim of the current work is to describe the behavior of one-electron probability density distribution and energy depending on the geometrical parameters of the system. Additionally we have studied the behavior of the interband quantum transitions both in the cases of single structure and the ensemble of the noninteracting QDs.

\section{Theory}

Consider one electron state in the cone-shaped quantum dot. One can present it as a spherical sector (see Fig 1). In this case the shape of quantum dot is different from the conical one because of the circularity of the base.However in the cases when the stretch angle of the spherical sector is small, the circularity of the base is negligible. Here the confinement potential both in radial and polar directions is chosen in a form of rectangular impenetrable walls:
\begin{figure}
    \centering
    \includegraphics[width=5cm]{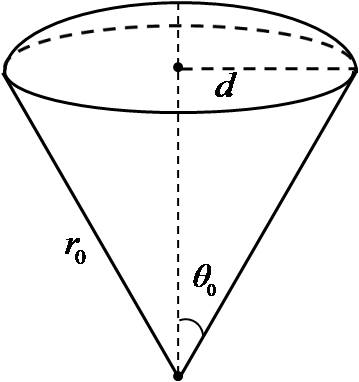}
    \caption{Quasi-conical quantum dot}
\end{figure}

\begin{equation}
V_{conf}^{rad}(r)=\begin{cases}
        0, &r<r_0\vspace{10pt}\\
        \infty, &r\geqslant r_0\\
    \end{cases},
\end{equation}
\begin{equation}
V_{conf}^{pol}(\theta)=\begin{cases}
        0, &\theta<\theta_0\vspace{10pt}\\
        \infty , &\theta\geqslant\theta_0\\
    \end{cases},
\end{equation}
where $r_0$ and $\theta_0$ are respectively the radius and stretch angle of the spherical sector (see Fig 1). Then in the spherical coordinates the Schrodinger equation of the system has a form
\begin{widetext}
\begin{equation}
-\frac{\hbar^2}{2\mu}\triangle_{r,\theta,\varphi}\psi(r,\theta,\varphi) + \left(V_{conf}^{rad}(r)+V_{conf}^{pol}(\theta)\right)\psi(r,\theta,\varphi)=E\psi(r,\theta,\varphi).
\end{equation}
\end{widetext}

The separation of variables is possible in this system, that is, one may seek the solution of the system in the form
\begin{equation}
\psi(r,\theta,\varphi)=R(r)P(\theta) \frac {e^{\imath m \varphi} }{\sqrt{2\pi}},
\end{equation}
where $P(\theta)$ and $R(r)$ are solutions of the following equations:
\begin{widetext}
\begin{equation}
\frac{d^2P(\theta)}{d\theta^2}+\cot\theta\frac{dP(\theta)}{d\theta}+\left( l(l+1)-\frac{m^2}{\sin^2\theta} \right)P(\theta)=0,
\end{equation}
\end{widetext}

\begin{equation}
\frac{d^2R(r)}{dr^2}+\frac{2}{r}\frac{dR(r)}{dr}+\left(k^2-\frac{l(l+1)}{r^2}\right)R(r)=0,
\end{equation}
where $k=\sqrt {\frac {2\mu E}{\hbar^2}}$ and $l$ are the analogues of the quantum numbers, describing the radial and polar motions of electron, respectively. Note that generally $l$ is not an integer value, as it is in case of motion in spherically symmetric fields, and depends on the boundary condition (2). The similar situation appears in case of quantum particle motion on the surface of the spherical segment (quantum ring on sphere) [27]. In this case the motion of electron is also restricted in polar direction. In the mentioned work [27] it was shown that the solution of angular equation (5) in the case of restricted motion is a linear combination of two solutions: (see also [28]):
\begin{widetext}
\begin{equation}
    \begin{aligned}
1)m\geq0& \\
&P(\theta)=C_1 \sin^m \frac{\theta}{2} \cos^m \frac{\theta}{2} {}_2 F_1  \left(m+l+1,m-l,1+m,\sin^2 \frac\theta 2\right)\\
&+ C_2 sin^m \frac{\theta}{2} cos^m \frac{\theta}{2} {}_2 F_1  \left(m+l+1,m-l,1+m,\cos^2 \frac\theta 2\right),\\
2)m<0&\\
&P(\theta)=C_1' \sin^{-m} \frac{\theta}{2} \cos^m \frac{\theta}{2} {}_2 F_1  \left(l+1,-l,1-m,\sin^2 \frac\theta 2\right)\\
&+ C_2' \sin^m \frac{\theta}{2} \cos^{-m} \frac{\theta}{2} {}_2 F_1  \left(l+1,-l,1-m,\cos^2 \frac\theta 2\right).\\
    \end{aligned}
\end{equation}
\end{widetext}
In our case the second solution diverges at $\theta=0$, so that it should be taken with coefficient zero. Finally, we have
\begin{widetext}
\begin{equation}
    \begin{aligned}
&m\geq0 \qquad P(\theta)=C \sin^m \frac{\theta}{2} \cos^m \frac{\theta}{2} {}_2 F_1  \left(m+l+1,m-l,1+m,\sin^2 \frac\theta 2\right),\\
&m<0 \qquad P(\theta)=C' \sin^{-m} \frac{\theta}{2} \cos^m \frac{\theta}{2} {}_2 F_1  \left(l+1,-l,1-m,\sin^2 \frac\theta 2\right).\\
    \end{aligned}
\end{equation}
\end{widetext}
The radial part of wave function is analogical to one appears in spherical impenetrable quantum well [3]:
\begin{equation}
R(r)=D \frac{1}{\sqrt{r}} J_{l+\frac{1}{2}}(kr).
\end{equation}
One can found the values of quantum numbers $l$ and $k$ from the implementation of the boundary conditions
\begin{equation}
P(\theta_0)=0,
\end{equation}
\begin{equation}
R(r_0)=0.
\end{equation}

The probability density distribution $|\psi|^2$ of different states in XOZ plane cross-section is shown in Fig. 2. Here the stretch angle of spherical sector is $\theta_0=30^\circ$. The Dirac notations $|klm\rangle$ are used, where the first number corresponds to the subscript of radial quantum number, the second one corresponds to the subscript of orbital quantum number, and the third one is magnetic quantum number. As it might be expected, the growth of quantum number $k$ when $l$ and $m$ are fixed brings to new peaks in radial direction (Fig. 2a - 2c). Respectively, the increase of quantum number $l$ leads to new peaks in polar direction (Fig. 2a - 2d - 2g). It should be noted, that due to the azimuthal symmetry of the problem all the states (expect the cases when $m=0$) are degenerate by the sign of the quantum number $m$. Besides that, in cases when $m\neq0$ the central peak in polar direction splits into two peaks (Fig. 2f, 2i, compare with Fig. 2a). The localization shift to the walls may be explained by the increase of motion intensity in azimuthal direction with the increase of quantum number $m$. In Fig. 2e, 2h the probability density distribution of the states $|220\rangle$ and $|221\rangle$ are presented.
\begin{figure}
    \centering
    \includegraphics[width=8cm]{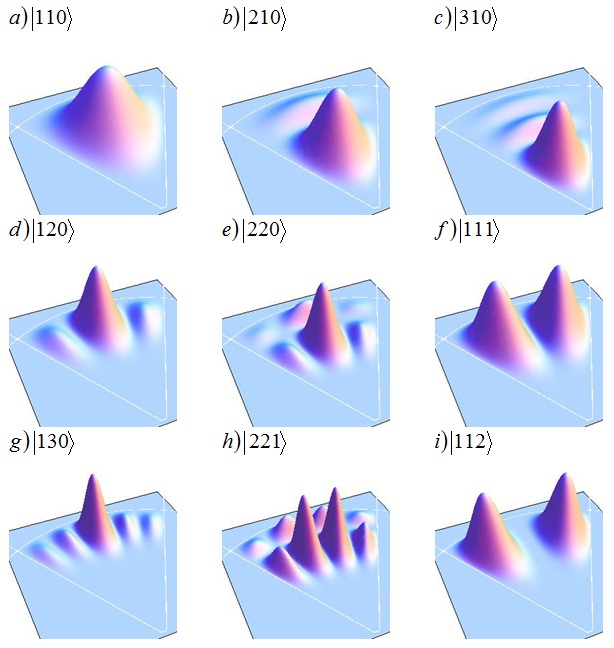}
    \caption{The probability density distribution of different states}
\end{figure}

It is interesting to discuss the energy spectrum dependence on the shape of quantum dot. Namely, in the Fig. 3 the dependence of the energy of different states on the stretch angle for fixed volume $V=\frac{2\pi}{3}r_0^3 (1-\cos\theta_0)=const$ of QD ($r_0=150{\AA}$ at $\theta_0=30^\circ$) is shown in dimensionless values $\varepsilon=E/\frac{\hbar^2}{2 \mu D^2}$, where $D=100{\AA}$. All calculations are performed for GaAs ($\mu=0.067m_0$). In the case of the fixed volume the increase of the stretch angle leads to the weakening of the size quantization in polar direction and to the strengthening in radial direction. Depending on the localization distribution, different states react in different ways on the combination of this two effects. Namely, in the Fig. 4 a, b the probability density distribution of $|130\rangle$ state is shown for different stretch angles. It is easy to see, that the weakening of size quantization in polar direction is much stronger than the strengthening in radial direction, which leads to the decrease of energy. However, it is not always easy to explain the behavior of different states via probability distribution because of competition of opposite effects in radial and polar directions.  Note that in case when $\theta_0=90^\circ$ the problem is reduced to the problem of electron states in semi-spherical quantum dot [29], and in case when $\theta_0=180^\circ$ one arrives to the problem of spherical quantum dot [3].
\begin{figure}
    \centering
    \includegraphics[width=8cm]{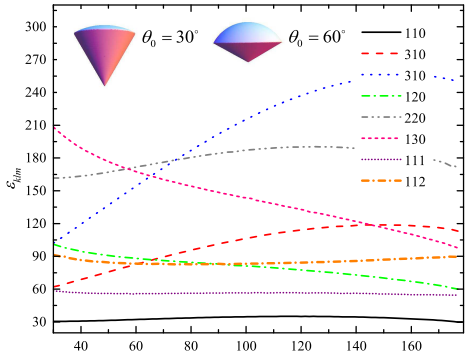}
    \caption{Dependence of electron energy spectrum on the quasi-conical QD stretch angle for the fixed volume ($r_0=150{\AA}$ at $\theta_0=30^\circ$)}
\end{figure}
\begin{figure}
    \centering
    \includegraphics[width=8cm]{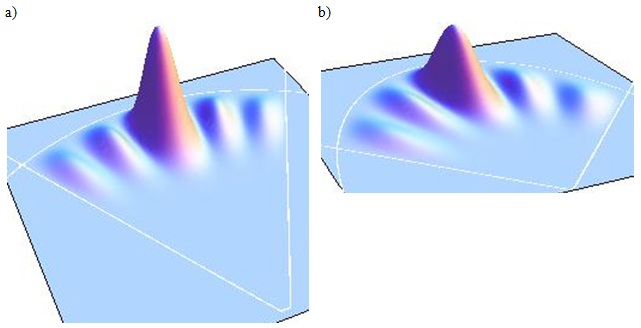}
    \caption{The probability density distribution of $|130\rangle$ state at stretch angles a) $\theta_0=30^\circ$, b) $\theta_0=60^\circ$}
\end{figure}

In Fig. 5 the dependence of the energy of different states on the stretch angle for fixed base $d=r_0 \sin\theta_0=const$ ($r_0=150{\AA}$ at $\theta_0=30^\circ$) is shown. One may easily explain the similar behavior of the all states on the basis of the dependence of volume on the stretch angle shown in Fig. 5 inset. It is a curve having a minimum, which corresponds to the maximums of the all states. In other words, in this case there is a strong dependence of localization area on the stretch angle, which causes the appropriate behavior of the all electron states.
\begin{figure}
    \centering
    \includegraphics[width=8cm]{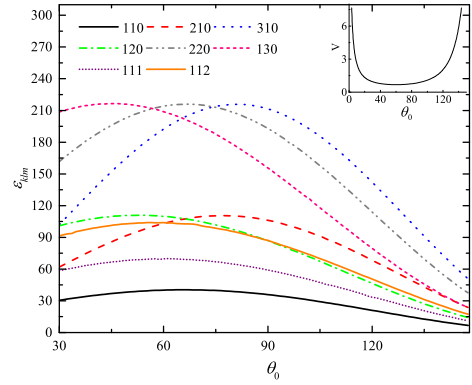}
    \caption{Dependence of electron energy spectrum on the quasi-conical QD stretch angle for the fixed base ($r_0=150{\AA}$ at $\theta_0=30^\circ$). In the inset the dependence of QD volume on the stretch angle for the fixed base is shown.}
\end{figure}

\section{Comparison}

As it was mentioned above, the problem of one-electron states in conical quantum dot was already discussed in several works [22], [25], [30, 31]. In the Ref. [25] the authors used a transformation of coordinates:
\begin{equation}
    \begin{cases}
        \frac{x}{z}=\rho \cos\varphi \\
        \frac{y}{z}=\rho \sin\varphi \\
        z=\omega
    \end{cases},
\end{equation}
which converts the conical domain into cylindrical shape (Fig 6 a):
\begin{equation}
    \begin{cases}
        0\leq \rho\leq \tan\theta_0\\
        0\leq\varphi\leq2\pi\\
        0\leq\omega\leq h
    \end{cases}.
\end{equation}

\begin{figure}
    \centering
    \includegraphics[width=8cm]{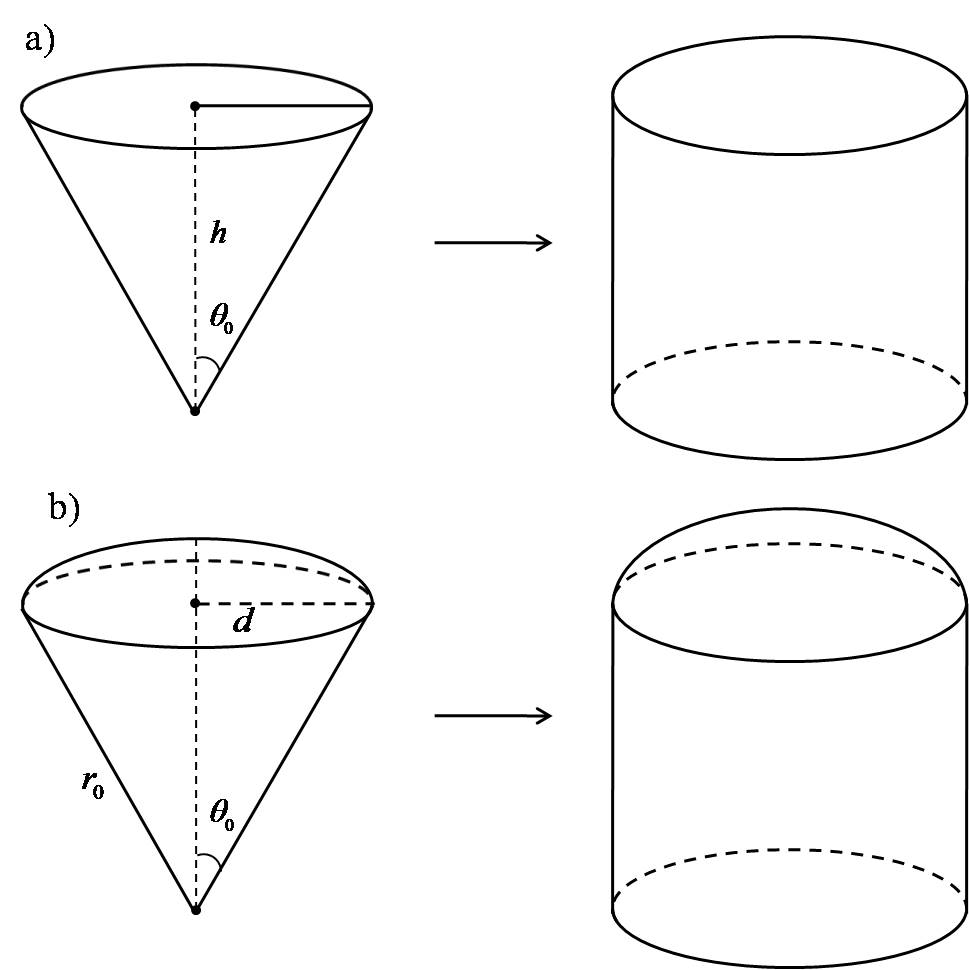}
    \caption{Visualization of mathematical transformation for the cases of a) conical QD, b) quasi-conical QD}
\end{figure}

Then for relatively low values of vertex angle $2 \theta_0$ in some approximation the separation of variables is possible. The same transformation in our case brings to the cylinder one of the bases of which is curved (Fig. 6 b):
\begin{equation}
    \begin{cases}
        0\leq \rho\leq \tan\theta_0\\
        0\leq\omega\sqrt{\rho^2+1}\leq r_0\\
    \end{cases}.
\end{equation}
Using the boundary conditions (24) of Ref. [22]
\begin{equation}
    \begin{cases}
        m=0,\pm1,\pm2,\ldots\\
        J_m\left(k_{\rho m i} \tan\theta_0\right)=0\\
        J_{\sqrt{k_{\rho i}^2-\frac34}}\left(k_{m i j} h\right)=0\\
    \end{cases},
\end{equation}
we have obtained the energy dependence on vertex angle for GaAs conical quantum dot.
\begin{figure}
    \centering
    \includegraphics[width=8cm]{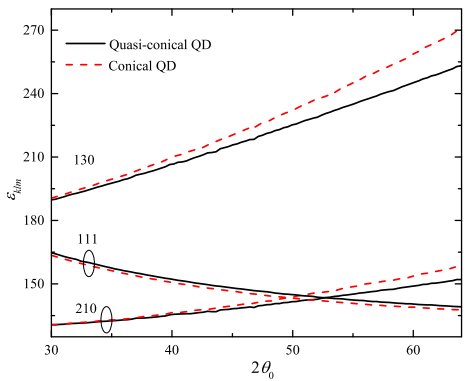}
    \caption{Dependence of electron energy spectrum on the vertex angle for the fixed volume ($h_0=90{\AA}$ at $\theta_0=30^\circ$ in Fig. 6a)}
\end{figure}

In the Fig. 7 the dependence of the energy of different states on the vertex angle for constant volume ($h_0=90{\AA}$ at $\theta_0=30^\circ$ in Fig. 6a) is shown. The solid lines correspond to our case, and the dashed lines are the energy values of conical quantum dot [25]. Such a good agreement of results for the all states should be explained by the fact that for relatively low values of vertex angle the base of quasi-conical quantum dot (our case) is approximately flat.

In the Fig. 8 the dependence of the energy of different states on the vertex angle for constant base ($h_0=90{\AA}$ at $\theta_0=30^\circ$ in Fig. 6a) is shown. The solid lines correspond to our case, and the dashed lines correspond to Ref. [25]. One can see the similar behavior of the all states. Some difference in values is explained by the fact that at the same values of the bases and vertex angles the localization area is greater in case of quasi-conical quantum dot.
\begin{figure}
    \centering
    \includegraphics[width=7.5cm]{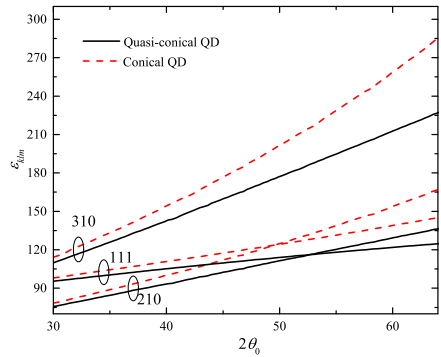}
    \caption{Dependence of electron energy spectrum on the vertex angle for the fixed base ($h_0=90{\AA}$ at $\theta_0=30^\circ$ in Fig. 6a)}
\end{figure}

\section{Quantum transitions}

As an application of the obtained results it is interesting to discuss interband quantum transitions in this system. The absorption coefficient in case of electron transition from valence band into conductive band is given by the formulae [3]:
\begin{widetext}
\begin{equation}
K(\omega, \theta_0, r_0)=A\sum_{m,m',l,l',k,k'}\left|\int \psi^e_{k,l,m}\psi^h_{k',l',m'}dv\right|^2\delta(\hbar\omega-E_g-E^e_{k,l,m}-E^h_{k',l',m'}),
\end{equation}
\end{widetext}
where $A$ is a quantity proportional to the square of the matrix element taken by Bloch functions [32], $\omega$ is the frequency of the incident light, $E_g$ is the band gap of bulk semiconductor. $\delta$ is the Dirac delta function, which provides the energy conservation law during the transitions. Here the selection rules for quantum numbers are $k'=k, l'=l, m'=-m$. Then the absorption coefficient takes a form
\begin{equation}
K(\omega, \theta_0, r_0)=A\sum_{m,l,k}\delta(\hbar\omega-E_g-\frac{h^2k^2}{2\mu_{red}}),
\end{equation}
where $\mu_{red}=\frac{\mu_h\mu_e}{\mu_h+\mu_e}$.

Thus, we obtained the absorption coefficient for a single structure having fixed stretch angle and radius. However in real physical problems we usually have an ensemble of QDs which differ from each other by their geometrical parameters. This is why it is necessary to find the absorption coefficient of similar structures. Here we assume that the interaction between QDs can be neglected, and that there is Gaussian distribution by the radiuses:
\begin{equation}
P(r_0)=\frac{1}{\sqrt{2\pi}\sigma} \exp \left[ -\frac{ (r_0- \bar{r_0}^2)}{2\sigma^2} \right].
\end{equation}
with a mean value of radius $\bar{r}_0$ and root-mean-square deviation $\sigma$. Then absorption coefficient takes a form
\begin{widetext}
\begin{equation}
K(\omega)=\int_{r_0^{'}}^{r_0^{"}} K(\omega, \theta_0, r_0) P(r_0) dr_0 = \frac{A}{\sqrt{2\pi}\sigma} \sum_{k,l,m} \frac{\exp \left[ -\frac{ (r_{00}- \bar{r_0}^2)}{2\sigma^2} \right]} {\left| f^{'}(r_{00} \right|},
\end{equation}
\end{widetext}
where $r_{00}$ is zero of the function
\begin{equation}
f(r_0)=\hbar\omega-E_g-\frac{\hbar^2 k^2(r_0)}{\mu_{red}}.
\end{equation}

In the Fig. 9 the dependence of the absorption coefficient on the incident radiation frequency is shown for different initial states. One can easily notice that in case of lower states the behavior of the absorption coefficient is peak-type. Instead, the more excited the state is, the more smooth the curve is. This can be  explained by the fact that more excited states are more sensitive to the changes of the geometrical parameters.
\begin{figure}
    \centering
    \includegraphics[width=8cm]{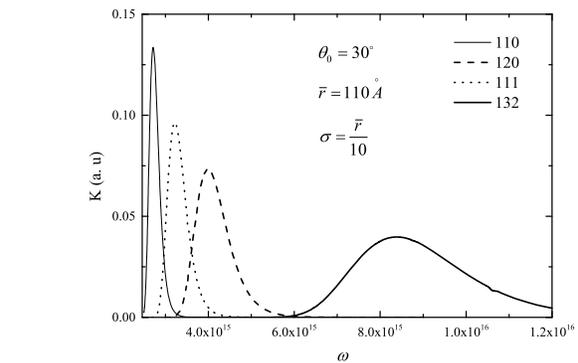}
    \caption{Dependence of the absorption coefficient (in arbitrary units) on the energy of incident radiation for different initial states}
\end{figure}

\section{Conclusion}

We have presented the exact solution of problem of one-electron states in quasi-conical quantum dot, which for relatively low values of stretch angle might be considered as a conical quantum dot. The probability density distribution for different states is presented. The energy spectrum of different states and its dependence on the parameters of quantum dot is obtained. It is shown that the results of the current work are in good agreement with the previously performed works. The quantum transitions caused by the absorption of incident radiation are discussed and the absorption coefficient is calculated.

\begin{acknowledgements}
The authors are grateful to Dr. Vram Mughnetsyan for discussion of the results and valuable comments. The work was performed within the framework of the State basic program "Investigation of physical properties of quantum nanostructures with a complex geometry and different confining potentials".
\end{acknowledgements}

\end{document}